\documentstyle[epsfig,twocolumn,prb,aps]{revtex}

\begin{document}
\draft

\twocolumn[\hsize\textwidth\columnwidth\hsize\csname @twocolumnfalse\endcsname

\title{
Field-induced gap in the spin-${1 \over 2}$ antiferromagnetic Heisenberg chain:\\
A density matrix renormalization group study}
\author{ Jizhong Lou$^{1}$, 
         Shaojin Qin$^{2,3}$, 
         Changfeng Chen$^{1}$
         Zhaobin Su$^{3}$
     and Lu Yu$^{4,3}$}
\address{
${}^{1}$Department of Physics, University of Nevada, Las Vegas, Nevada 89154 \\
${}^{2}$Department of Physics, Kyushu University, Hakozaki, Higashi-ku,
        Fukuoka 812-8581, Japan \\
${}^{3}$Institute of Theoretical Physics, P O Box 2735, Beijing 100080,
        P R China\\
${}^{4}$ Abdus Salam International Center for Theoretical Physics, P. O. Box 586,
        34100 Trieste, Italy\\
}
\date{ \today }
\maketitle

\begin{abstract}
We study the spin-${1 \over 2}$ antiferromagnetic Heisenberg chain in both 
 uniform  and (perpendicular) staggered magnetic fields using the 
density-matrix renormalization-group method.  This model has been shown 
earlier to describe the physics of the copper benzoate materials in magnetic field.  
In the present work, we extend the study to more general case for a systematic 
investigation of the field-induced gap and  related properties of the
spin-${1 \over 2}$ antiferromagnetic Heisenberg chain.  In particular, we explore 
the high magnetic field regime where interesting behaviors in the field-induced gap, 
magnetization, and spin correlation functions are  found.
Careful  examination of the low energy properties and  magnetization reveals interesting
competing effects of the staggered and  uniform fields.  The incommensurate behavior in
the spin correlation functions is demonstrated and discussed in detail. The present
work reproduces earlier results in good agreement with experimental data on copper
benzoate and predicts new interesting field-induced features at very high magnetic field.
\end{abstract}


]

\section{Introduction}
\label{intro}

The quasi-one-dimensional magnetic materials have attracted considerable
experimental and theoretical interest since Haldane's pioneering work \cite{haldane83} 
that pointed out the difference between the integer spin Heisenberg 
antiferromagnetic (HAFM) chains and the half-integer spin chains.  By mapping
the Heisenberg spin chains onto the $O(3)$ nonlinear sigma 
model,\cite{affleck89} Haldane conjectured that the low-energy excitation 
spectrum displays a finite energy gap for the integer spin systems while 
for half-integer spin chains it is gapless.  These conjectures have been verified by
later detailed studies.

In the linear chain HAFM family, the spin-${1 \over 2}$ chain is of particular interest
since most of its properties can be obtained exactly. These exact results serve 
as benchmarks in testing the validity of various approximation schemes.  
The low energy excitation spectrum in this system has no gap; 
the elementary excitations are spin-${1 \over 2}$ spinons; the ground state is 
quasi-long-range ordered, and the spin-spin correlations display power law decay.  

The effect of an applied magnetic field on the spin-${1 \over 2}$ Heisenberg chain has 
also been studied to gain more insight into the physics of such a system.  When  
the external  magnetic field is present, the Hamiltonian of the system is written as
\begin{equation}
H=J \sum_i ( {\bf S}_i \cdot {\bf S}_{i+1} - h_u S^z_i ),
\label{magham}
\end{equation}
where $J$ is the coupling constant, ${\bf S}_i$ the spin-${1 \over 2}$ operator 
on site $i$, $S^{z}_i$ the $z$ component of ${\bf S}_i$, $h_u = g \mu_B H/J$ is the 
effective dimensionless uniform field, $g$ the average effective gyromagnetic ratio 
and $H$ the applied magnetic field.  
When $h_u =0$, the critical  wave vector of the gapless excitation is located  at  0 and $\pi$.
The applied field will shift the critical wave vector of the gapless excitation away from 
0 ( for transverse spin correlations)  and $\pi$   ( for longitudinal spin  correlations)
to  incommensurate values while the excitation remains gapless
until the field is larger than its saturation value $h_u=2.0$.\cite{muller81}
When $h_u=2.0$, the magnetization is saturated with all  spins on the lattice sites 
oriented parallel to the applied uniform field.  Further increase in the applied field 
leads to the opening of a gap in the low energy  spectrum with its magnitude 
changing almost linearly with the applied field, corresponding to flipping one spin 
to its opposite direction.

The field-induced incommensurate state was first observed in the 
neutron scattering measurements on copper benzoate, 
Cu(C$_6$D$_5$COO)$_2$$\cdot$3D$_2$O.\cite{dender97} 
Copper benzoate is a linear chain spin-${1 \over 2}$ AFM \cite{date70}
with  coupling constant $J \sim$ 1.57 meV.\cite{dender96} 
In this material, the effective spin-${1 \over 2}$ Cu$^{2+}$ ions form a linear
chain structure, but the two neighboring copper sites are not totally equivalent.
Because of the small value of the coupling constant $J$, it is possible to
study the high field (large $h_u$) properties of Hamiltonian (\ref{magham}) 
and investigate how the induced incommensurate soft mode behaves with the 
changing magnetic field. 

In the copper benzoate experiment, \cite{dender97} 
in addition to the field-dependent incommensurate low energy modes, 
an unexpected non-zero energy gap induced by the magnetic field was 
detected. The value of the gap varies with the magnitude and the
relative orientation of the applied magnetic field.\cite{dender97}
It was first suggested\cite{dender97}  that the unexpected gap is caused by the 
inequivalence of the Cu sites leading to an effective staggered $g$ tensor,
which in turn gives rise to an effective staggered field in addition to the applied uniform
magnetic field. Later, detailed analysis \cite{oshikawa97} shows that an additional staggered
Dzyaloshinskii-Moriya\cite{dm1,dm2} interaction term provides similar contribution
along with the staggered $g$ tensor, and both have the same order in magnitude. 
The Dzyaloshinskii-Moriya interaction together with the staggered $g$ factor can account
for the observed non-zero energy gap in copper benzoate.
After making some assumptions and ignoring the small exchange anisotropy,
the effective Hamiltonian to describe the 
copper benzoate in  the  magnetic field   can be written as:\cite{oshikawa97}
\begin{equation}
H=J \sum_i [ {\bf S}_i \cdot {\bf S}_{i+1} - h_u S^z_i - (-1)^i h_s S^x_i ],
\label{ham}
\end{equation}
where $h_s$ is the induced effective dimensionless staggered field, 
which is the key term to account for the observed non-zero energy gap. 
The magnitude of $h_s$ depends on the magnitude and relative
direction  of the applied uniform field $h_u$ with respect to the sample.

Hamiltonian (\ref{ham}) has been studied using the bosonization approach,
mapping on the Sine-Gordon model, and form-factor techniques. 
The gap and magnetization behavior,\cite{oshikawa97,affleck99}
the dynamical magnetic susceptibility,\cite{essler98,affleck99} 
the specific heat \cite{essler99} and the electron spin resonance\cite{oshikawa99} experiments
have been analyzed using the sine-Gordon quantum field theory.
There have also been   numerical studies of the excitation energy
and transition amplitudes in the staggered magnetic field, based on the Bethe Ansatz
solutions.\cite{fled98}
These studies focus mainly on the parameter range   corresponding to
the reported copper benzoate magnetic field experiment. \cite{dender97}
The field-induced gap and related magnetization, as well as   spin correlations  in 
generic spin-${1 \over 2}$ antiferromagnetic Heisenberg chains are
of great interest but yet to be fully understood.  Of particular interest
is the study of a wider range of parameter  and very high magnetic
field conditions beyond those probed by the experiment and previous theoretical work,
which may prove to be of significant importance in understanding 
the entire range of field-induced phenomena and the underlying physics in this 
interesting system.

In this paper, we report  results of our numerical calculations of the ground 
state and the low energy excitations of Hamiltonian (\ref{ham}) using the 
density matrix renormalization group (DMRG) method.\cite{dmrg} 
We study the most general case and take $h_u$ and $h_s$ as independent variables 
in our calculations.  We find that the critical (saturation) uniform field
$h_u^c=2.0$ serves as an important reference point in understanding the obtained 
results.  When fixed $h_u$ is lower than or equal to $h_u^c$, the induced energy
gap increases with $h_s$ as a power  with exponent $\sim \frac{2}{3}$,   
and for fixed $h_u$ larger than $h_u^c$, the small $h_s$ dependence
is exponential. On the other hand, when $h_s$ is fixed, the $h_u$ dependence of
the gap displays a minimum around $h_u^c$.  When $h_u$ and $h_s$ increase simultaneously
with a fixed ratio, the gap increases with the field when the ratio is small but
develops a minimum around $h_u^c$ at larger ratios.  The magnetization 
results are consistent with the intuitive expectation in general.  The most interesting 
features are obtained for  fixed $h_u < h_u^c$ and in small fixed $h_s$ 
cases.  When $h_u$ is lower than $h_u^c$ and fixed, the existence of a
small staggered field enhances the uniform magnetization
instead of suppressing it; similar effects are also observed
for small fixed $h_s$ case, where the staggered magnetization
increases until $h_u$ approaches $h_u^c$.  The uniform field induced incommensurate 
behavior is also studied, and the results show competing effects
of the uniform and  staggered fields, with the staggered field frustrating the 
incommensurate state.

In the following, the numerical DMRG results will be presented in Section \ref{sec2} 
and a summary  given in Section \ref{sec3}.

\section{DMRG Results}
\label{sec2}

The DMRG method \cite{dmrg} is a powerful tool for the calculation of low lying 
states of quasi-one-dimensional systems and has been developed to calculate 
other properties of many strongly correlated systems.\cite{conf}
The accuracy of the DMRG calculations on spin chains is generally high.  This has
also been verified in the S=1 (Ref. \onlinecite{spinone})
and S=2 (Ref. \onlinecite{spintwo}) Heisenberg chain calculations.
In the present work, we employ the periodic boundary conditions (PBC) and use the
infinite chain length algorithm of DMRG. We retain as many as
500 optimal states and compute up to chain length N=100 in 
each calculation. The largest truncation errors are of the order
of 10$^{-9}$ for zero uniform field calculations and 10$^{-6}$
for non-zero uniform field calculations. To simplify the discussion,
we set the coupling constant $J$ as the energy unit, $h_u$ as the effective
uniform field and $h_s$ as the effective staggered field.

For the Hamiltonian considered here, the effect of the uniform field
is to induce a uniform magnetization and shift the critical wave vector
from $\pi$ (In this paper we will concentrate on this case for the pitch vector,
where a peak in the static structure factor is expected), 
and the staggered field will induce a non-zero energy gap 
between the ground state and the lowest excited state.  A non-zero staggered 
magnetization will also be expected for finite staggered fields. The nature 
of the ground state depends on the competition of the uniform and 
the staggered field.  We will discuss the energy spectrum of the Hamiltonian first, 
and then show the ground-state magnetization behavior and investigate 
the incommensurate behavior of the spin correlation functions.

\subsection{Energy Gap}

When  the staggered field is present alone in the spin-${1 \over 2}$ chain,
that is, $h_u$=0 in Hamiltonian (\ref{ham}), 
\begin{equation}
H=\sum_i [ {\bf S}_i \cdot {\bf S}_{i+1} - (-1)^i h_s S^x_i ]
\label{stagham}
\end{equation}
the x-component of the total spin $S^x_{tot}$ is conserved.
This property can be used in DMRG calculations to reduce
the dimension of the Hilbert space  to be considered.

The spin-1 case of Hamiltonian (\ref{stagham})
has been studied in detail.\cite{lou99} For the standard
spin-1  Heisenberg chain, the low energy  spectrum is gapful, and 
the lowest excited state is a spin triplet known as Haldane triplet. 
The presence of a staggered field will split the Haldane triplet into two branches, 
the transverse branch and the the longitudinal branch.
Both branches are gapful and the gap increases with the staggered field. 

For the spin-${1 \over 2}$ case, the excitation spectrum for zero-field  chain 
is gapless, and the presence  of the
staggered field may also open a finite gap between the ground state
and the low-energy continuum.  It is expected that
the gapful  excitations  will also split into two branches, with the x-component of 
total spin  $S^x_{tot}$=0 (longitudinal branch) and 1 (transverse branch).  

\begin{figure}[ht]
\epsfxsize=3.3 in\centerline{\epsffile{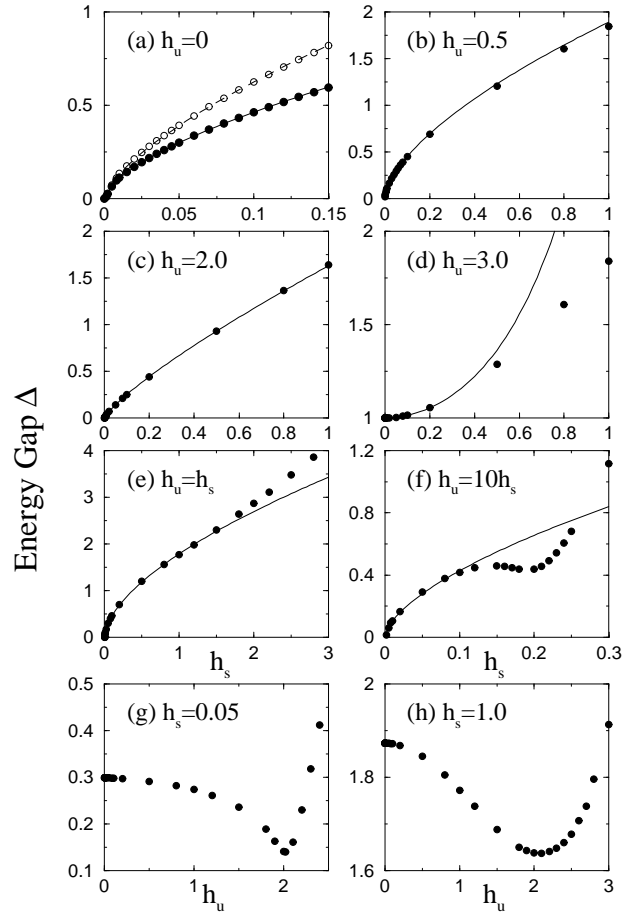}}
\caption[]{ Field dependence of the induced energy gap for
Hamiltonian (\ref{ham}) with different $h_u$ and $h_s$
relations. The DMRG results are shown by circle, the lines
denote the fitting curves.  In (a), the filled circles denote the transverse 
branch,  while the empty circles the longitudinal branch.
We note that the "y" axis in (d) starts from $\Delta=1$ which is
different with these starts from $\Delta=0$ in others.}
\label{fig1}
\end{figure}

The field dependence of the gap of the longitudinal 
and  transverse branches for Hamiltonian (\ref{stagham})
is shown in Fig. \ref{fig1} (a).
It is clear that the low energy spectrum becomes gapful as soon as 
the staggered field is present. The magnitude of both 
 longitudinal and   transverse modes increases 
when the staggered field becomes larger with the longitudinal mode goes up
faster than the transverse one.  This behavior is exactly the same as for  the spin-1 chain.
But for the spin-1 chain in  the staggered field, the increase of
the longitudinal gap is about three (two) times faster than the transverse one
for small (large) staggered field.  For the spin-${1 \over 2}$ chain case, this ratio 
is smaller. The two gaps are fitted using the equation
\begin{equation}
\Delta = a h_s^b,
\label{gapcurve}
\end{equation}
where $a$ and $b$ are fitting parameters. The least square fitting gives:
\begin{eqnarray}
\Delta_L & = & a_L h_s^{b_L} =  2.97 h_s^{0.678}, \nonumber \\
\Delta_T & = & a_T h_s^{b_T} =  1.97 h_s^{0.63},
\label{fitgap0}
\end{eqnarray}
where $\Delta_L$ and $\Delta_T$ denote the longitudinal  and 
transverse gap, respectively. 
The ratio of the gap increase coefficient 
of the longitudinal mode to the transverse mode
$a_L /a_T$ is about 1.5. 
The fitting curves Eq. (\ref{fitgap0})
are also shown in Fig. \ref{fig1}(a). 
For larger staggered field, the fitting is almost perfect, 
while it is  not very good for very small staggered field.
(This deviation is not visible in Fig. \ref{fig1} (a).)
It should be noted here that the numerical error of DMRG results for
small staggered field is much bigger than that for large field.

When the uniform field in Hamiltonian (\ref{ham}) is not vanishing,
the competition of the uniform    and the staggered field
needs to be taken into account.  The presence of the uniform field will affect 
the behavior of the energy gap.
In the non-zero uniform field case, the calculation is more difficult, 
since now even $S^x_{tot}$ is not conserved any more.  There is no good
quantum number which can be used to reduce the relevant Hilbert space dimension.
To investigate the gap behavior, 
the ground state and the lowest excited state must be calculated 
at the same time, and at least two states must be targeted in each calculation.

To study the effect of the uniform field on the gap behavior,
we first calculate the staggered field dependence of the gap
when uniform field $h_u$ is a non-zero constant.
As described above, when  the uniform field
is present alone, there is a critical point $h_u^c=2.0$, 
where the magnetization is saturated.
We need to perform calculations at different values of $h_u$,  
one in each part of the $h_u$ phase space.
We choose $h_u$=0.5 for the small uniform field case and $h_u$=3.0
for the large field case. We also calculate the $h_s$ dependence
of the gap at the uniform field critical point $h_u^c$=2.0.
   
The results for $h_u$=0.5, 2.0 and 3.0 are shown in 
Fig. \ref{fig1} (b), (c), (d),  respectively.
For $h_u$=0.5 and 2.0, the spectrum is gapless when $h_s$=0, 
and an energy gap opens up when the staggered field is present. 
The gap increases with the staggered field $h_s$ following the
same function Eq. (\ref{gapcurve}) as in the $h_u$=0 case, although
with different parameters.  Least square fitting gives
$a$=1.89, $b$=0.624 for $h_u$=0.5 and $a$=1.63, $b$=0.81 for $h_u$=2.0.
Comparison with  results in Fig. 1 (a) shows that both the coefficient and the
exponent in the $h_u=0.5$ case differ only slightly from those of the transverse
branch for $h_u$=0; for $h_u$=2.0, the difference becomes more pronounced.
The calculated results show that in the uniform field gapless phase ($h_u \le 2.0$),
the induced gap is affected by the uniform field
only when the uniform field is strong enough.
For $h_u$=3.0, the gap dependence is different. 
The energy spectrum is gapful even when $h_s$=0.  With the application of the 
staggered field, the gap increases.
The increase of the gap is nearly exponential when the staggered
field is not very strong, following
$\Delta=\exp(a h_s^b)$ with  parameters a=1.186 and b=1.939. 
When the staggered field is large enough, the gap increase 
deviates from the exponential behavior.

In the experiment  on Cu benzoate,\cite{dender97} the magnitude of the induced
staggered field  depends  on the applied uniform field. Roughly
speaking, when the relative orientation of the applied field is fixed, the
induced staggered field increases linearly with the applied uniform field. 
To compare with experiment directly, we have
also considered the case when $h_s$ and $h_u$ increase at the
same time with the ratio $h_s/h_u$ being fixed. In Fig. \ref{fig1} (e) and (f), 
we present the calculated results of the energy gap with 
the staggered field for the cases of $h_u = h_s$ and $h_u = 10 h_s$.   
In both cases, the field dependence of the
gap for smaller staggered field can also be fitted with
function  (\ref{gapcurve}).
For $h_u = h_s$, the fitting holds at $h_s=h_u < 2.0$
with $a$=1.786 and $b$=0.594, while
for $h_u = 10 h_s$, it holds for $h_s < 0.1$ with $a$=1.755 and $b$=0.613.
The gap for $h_u=h_s$ case increases monotonically but 
that for $h_u=10 h_s$ exhibits a minimum near $h_s \sim 0.2$ ($h_u \sim 2.0$),
and then increases rapidly.  The  largest  induced 
staggered field produced in experiment\cite{dender97} is  about $h_s \sim$ 0.05, 
and it is not big enough to detect the gap minimum 
shown in Fig. \ref{fig1} (f).  Experiments at higher magnetic field are needed to
test this predicted phenomenon.  Here we can see again that the critical point
$h_u^c =2.0$ plays an important role in separating different regions where the
scaling behavior of the field-induced gap shows qualitative difference.

The effect of the uniform field on the spin gap induced by the
staggered field can be studied directly by calculating the uniform field 
dependence of the gap at fixed staggered field.
From the above results, we have learned that the system stays
in one single gapful phase when  the staggered field is present alone,
so we just choose two sets of $h_s$, $h_s$=0.05 as the small staggered field
limit and $h_s$=1.0 as the large staggered field limit.
In Fig. \ref{fig1} (g) and (h), we show the gap behavior 
with the change of the uniform field for the two cases. 
It is clearly seen that in both cases
the gap decreases when the uniform field increases from
zero and reaches its minimum near $h_u \sim 2.0$,
followed by a rapid increase with further increase in the uniform field.
The increase of the gap after the minimum is approximately linear for both cases.

Detailed calculations around the gap minimum provide more information on the 
gap behavior.  For $h_s=0$, the system is gapless until the
uniform field reaches the saturation point $h_u^c =2.0$.
When $h_s$ is not zero, the system has a finite energy gap,
and the presence of a small uniform field may suppress the gap.
However, for small uniform fields, the suppression of the gap 
is negligible; it  becomes visible only when the uniform field is large enough. 
The gap minimum for non-zero $h_s$ occurs when the uniform field
$h_u$ is slightly larger than 2.0.  It occurs at $h_u \sim 2.02$ for 
$h_s$=0.05 and $h_u \sim 2.1$ for $h_s$=1.0.
This means that the minimum gap uniform field value $h_u^{min}$
increases slowly with $h_s$.
After the gap minimum, its behavior is dominated by the uniform field, 
and the gap increases almost linearly with $h_u$. 
We should emphasize here that the minimum gap for both cases is not
zero, and for the $h_s$=1.0 case the decrease in magnitude is a little
bigger than that for the $h_s$=0.05 case.

\subsection{Magnetization}

The existence of a non-zero uniform field will induce a magnetization
in the system, and the existence of a non-zero staggered field will induce an
additional staggered magnetization.
The staggered magnetization $M_s (N)$ and the uniform magnetization
$M_u (N)$ of the system with a finite chain length $N$ are defined as
\begin{eqnarray}
M_s (N) &=& \frac{1}{N} \sum_i (-1)^i \langle S^x_i \rangle, \nonumber \\
M_u (N) &=& \frac{1}{N} \sum_i \langle S^z_i \rangle. 
\label{finitemag}
\end{eqnarray}

\begin{figure}[ht]
\epsfxsize=3.3 in\centerline{\epsffile{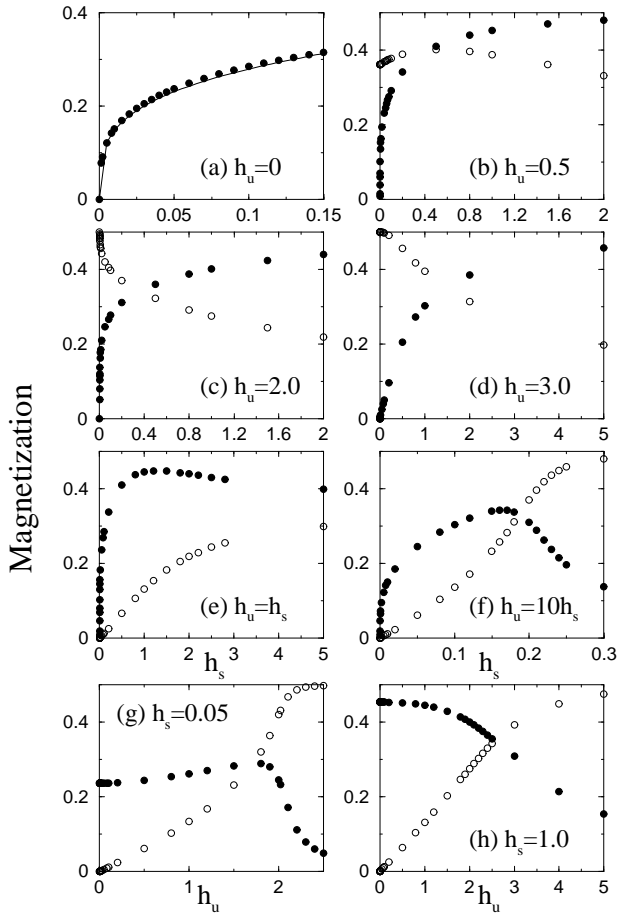}}
\caption[]{ Magnetization curves for eight parameter sets 
of $h_u$ and $h_s$. The filled (empty) circles are for the
staggered (uniform) magnetization.  The solid line in (a) is the 
fitting curve described in text.  For the $h_u$=0.5 case in (b), the uniform 
magnetization is amplified 6 times to bring it up to the same scale with the
other panels.}
\label{fig2}
\end{figure}

The results for the thermodynamic limit
\begin{eqnarray}
M_s &=& \lim_{N \to \infty} M_s (N) ,\nonumber \\
M_u &=& \lim_{N \to \infty} M_u (N).\label{mag}
\end{eqnarray}
can be obtained by studying different chain-length systems.
The results of the magnetization and the staggered magnetization
for the eight parameter sets ($h_u$, $h_s$) used in the previous subsection 
are shown in Fig. \ref{fig2}.

When the uniform magnetization is absent ($h_u = 0$),
the induced staggered magnetization increases with the
staggered field and approaches the saturation value 0.5 
when $h_s \rightarrow \infty$.
From the field dependence of the staggered magnetization,
we can extract the staggered magnetic susceptibility
$\chi^{(s)} = \frac{\partial M_s}{\partial h_s}$.
In Fig. \ref{fig2} (a), we can see clearly that the staggered 
magnetic susceptibility goes to infinity when
the staggered field $h_s \rightarrow 0$  instead of
approaching  a constant as in the case of the spin-1 chain. This is because the
spin-${1 \over 2}$ chain is gapless for zero staggered field while it
has a finite gap in the spin-1 case.
The magnetization curves are fitted using the following function:
\begin{equation}
M_s = a h_s^b
\label{magfit}
\end{equation}
with $a$=0.527 and $b$=0.277. The fitting line is also shown 
in Fig. \ref{fig2} (a). In the $h_s$ range shown in Fig. \ref{fig2} (a),
the fitting is good, but it should be
noted that the fitting Eq. (\ref{magfit}) will not be valid for very
large $h_s$, since it diverges when $h_s \rightarrow \infty$ 
instead of approaching  the finite value 1/2 which is the strong
staggered field limit of Hamiltonian (\ref{stagham}). 

When $h_u$ is finite and fixed, the staggered magnetization 
increases monotonically with the staggered field 
and approaches the saturation value $M_s$=0.5 when $h_s$ goes to infinity,
the same as in the $h_u=0$ case.
The magnetizations for $h_u=$0.5, 2.0, and 3.0 are shown
in Fig. \ref{fig2} (b), (c), and (d), respectively.
Because the zero staggered field energy spectrum for $h_u$=0.5 and
$h_u$=2.0 is gapless, the zero-field staggered magnetic susceptibility
at $h_s$=0 is still divergent. For $h_u$=3.0, the $h_s=0$ system is gapful,
so the zero-field staggered magnetic susceptibility has a finite value, 
$\chi^{(s)} (0)$=0.299.  For $h_u$=2.0 and 3.0, 
the uniform magnetization is saturated when the staggered field is absent;
it decreases from the saturation value 0.5 when a staggered field is applied. 
When the staggered field is weak, the way the magnetization changes is different
for the two cases.  It decreases rapidly for $h_u$=2.0 but slowly for $h_u$=3.0. 
For $h_u$=0.5, at zero staggered field, the uniform magnetization has a finite
value but is not saturated; when the staggered field increases, 
the uniform magnetization will also increase and reach a maximum
when $h_s \sim$ 0.5 before decreasing with further increasing staggered field.  

In Fig. \ref{fig2} (e) and (f), we show the field dependence of the
magnetization for the $h_u=h_s$ and $h_u= 10 h_s$ cases, respectively.
When the uniform and staggered fields increase simultaneously from zero,
both  staggered  and  uniform magnetizations increase. 
For both cases, the zero-field staggered susceptibility is divergent
and  the zero-field uniform susceptibility is zero, we should note
here that the zero-field uniform susceptibility is not shown clearly
in Fig. \ref{fig2} (e) and (f), its zero value can only be obtained
when the results for very small field is investigated.  When $h_u=h_s$, the 
effect of the staggered field rises rapidly and is dominating at small fields.
When the field increases, the effect of the uniform field becomes more important and
must be taken into account. At $h_s = h_u \sim$ 1.2, the staggered 
magnetization reaches its maximum, and it decreases with further increasing field.
The uniform magnetization increases monotonically
with the increasing fields. Both  staggered  and
 uniform magnetizations approach finite but nonsaturated values
when the field goes to infinity.
For $h_u=10 h_s$, the effect of the staggered field is also dominating
at small field. But at large field, the effect of the uniform field 
becomes dominant. The staggered magnetization decays fast 
to a small finite value $\sim$ 0.05
after reaching its maximum. The uniform magnetization increases monotonically
and approaches a nearly saturated value 0.498 at high enough field.

When the staggered field is fixed, at $h_u=0$ the staggered 
magnetization is finite while the uniform magnetization is zero.
The uniform field dependence of the magnetization
for $h_s$=0.05 and $h_s$=1.0 is shown in Fig. \ref{fig2}(g) and (h).
For $h_s$=0.05, the staggered magnetization
increases with the uniform field first, reaches its maximum at $h_u \sim$ 1.8, 
and then decays to zero rapidly.
For $h_s$=1.0, the  staggered magnetization decreases with the
uniform field monotonically. But the decrease is not rapid for small $h_u$.
In both cases, the uniform magnetization increases with the uniform field.
At small uniform field, the uniform magnetization increases linearly
with the uniform field.  We have obtained zero-field uniform
magnetic susceptibility $\chi^{(u)}(0)$ = 0.1185 for $h_s$=0.05
and $\chi^{(u)}(0) $ = 0.1283 for $h_s$=1.0.

For the $h_u$=0.5 ($h_s$=0.05) case, the existence of a small 
staggered (uniform) field enhances the corresponding uniform (staggered)
magnetization instead of suppressing it.
This phenomenon can be explained intuitively.
In these cases, when a small uniform (staggered) field
is applied, the coupling between neighboring spins
is weakened. While this uniform (staggered) field
is not strong enough to destroy the effect of the
stronger staggered (uniform) field, it enhances
the ratio between the effective uniform (staggered) field
and the effective coupling constant and, consequently, the induced uniform 
(staggered) magnetization.

\subsection{Correlation Function and Incommensurate Behavior}

We define three ground-state spin correlation functions for chain length $L$,
(i) ${\cal C}_u$ parallel to the uniform magnetic field, (ii) ${\cal C}_s$
parallel to the staggered field, and (iii) ${\cal C}_y$ along the remaining ($y$)
axis, as
\begin{eqnarray}
{\cal C}_u(i-j) &=& \langle S^z_i S^z_j \rangle, \nonumber \\
{\cal C}_s(i-j) &=& \langle S^x_i S^x_j \rangle, \nonumber \\
{\cal C}_y(i-j) &=& \langle S^y_i S^y_j \rangle.
\label{corr}
\end{eqnarray}
The correlation function ${\cal C}_y$ is expected to  display exponential decay
because of the existence of the spin gap induced by the staggered field.
${\cal C}_s$ and ${\cal C}_u$ do not decay exponentially because
of the effects of the non-zero staggered   and 
uniform magnetization, respectively.  ${\cal C}_u$ is  also expected to
show incommensurate behavior 
due to the existence of the uniform field. 
These correlation functions have the following form:\cite{white96}
\begin{eqnarray}
{\cal C}_u (l) & = & M_u^2+
(-1)^l A_1 \frac{e^{-l/\xi}}{\sqrt{l}} \cos(a l + \theta_0),
\nonumber \\
{\cal C}_s (l) & = & (-1)^l M_s^2+ (-1)^l A_2 \frac{e^{-l/\xi}}{\sqrt{l}},
\nonumber \\
{\cal C}_y (l) & = & (-1)^l A_3 \frac{e^{-l/\xi}}{\sqrt{l}},
\label{corr1}
\end{eqnarray}
where 
$M_u$ and $M_s$ are the uniform and staggered magnetization, respectively,
$A_1$, $A_2$, $A_3$, $a$, $\theta_0$ are $l$ independent
constants, and $\xi$ is the correlation length.  It should be noted 
here that the length $\xi$ may not be the same along   three different directions.

\begin{figure}[ht]
\epsfxsize=2.0 in\centerline{\epsffile{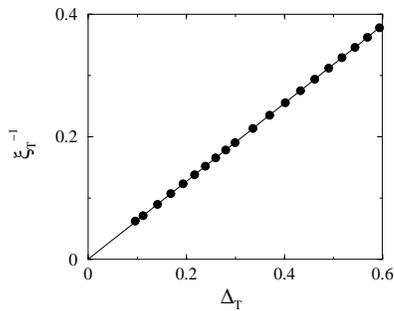}}
\caption[]{ The  inverse correlation length
for the transverse branch of the zero uniform field
case vs the transverse gap at different staggered fields.
The solid fitting line is $\xi^{-1}_T = 0.6364 \Delta_T$.}
\label{fig3}
\end{figure}

From the spin correlation functions Eq. (\ref{corr1}),
we can extract the corresponding correlation length. 
When $h_u$=0, the obtained correlation length is  a function of the staggered field.
The product of the correlation length and the spin gap gives the
spin wave velocity of the system. In Fig. \ref{fig3}, we show
the relation between the inverse correlation length and the transverse gap.
It is seen that the curve goes linearly which means that the spin wave
velocity does not change with the magnitude of the staggered field.
The linear fitting of the line gives the spin wave velocity 
$v =  \Delta_T \xi_T = 1/0.6364 = 1.5713$, in good
agreement with the exact spin wave velocity for the  spin-${1 \over 2}$ chain
$\frac{\pi}{2} \approx$ 1.5708. As in the spin-1 chain case,\cite{lou99} 
this also serves as an independent check of the self-consistency of our calculations.
From the figure, we can see for zero stagger field spin-${1 \over 2}$ chain,
the spectrum is gapless, $\Delta=0$, so the correlation length is infinite.
For other cases in our calculations, the correlation length is
difficult to obtain, because the chain length is short and the 
numerical error is bigger for those cases. But we can conclude
from our results that in general the spin wave velocity is not
a constant any more, and,  instead,  it changes with the applied uniform field.

\begin{figure}[ht]
\epsfxsize=2.0 in\centerline{\epsffile{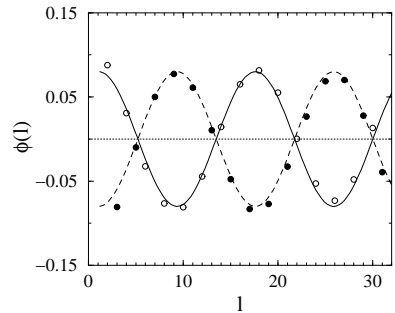}}
\caption[]{Incommensurate behavior in
$\phi(l) = \sqrt{l} e^{l/\xi} ({\cal C}_u(l) - M_u^2)$ 
for $h_u$=0.5, $h_s$=0.05 and chain length L=100 system. 
The correlation length $\xi \approx 4.05$.
The odd and even $l$ are denoted by filled and empty circles,
respectively.
The solid and dashed fitting lines are $\pm 0.08 \cos (0.38 l -0.407)$.}
\label{fig4}
\end{figure}

The cosine function in ${\cal C}_u$ comes from the incommensurate 
behavior induced by the uniform field.  Because of the existence of the staggered field, 
the net incommensurate behavior is a result of  competition between
 $h_u$ and $h_s$.  In Fig. \ref{fig4},
we  present $\phi(l)=\sqrt{l} e^{l/\xi} ({\cal C}_u(l) -M_u^2)$ 
at chain length 100
as a function of $l$ for $h_u$=0.5 and $h_s$=0.05, where $M_u$ is
obtained by the magnetization calculation discussed above. 
This clearly shows the existence of incommensurability in the system.
In this case, the correlation length $\xi \approx$ 4.05.

\begin{figure}[ht]
\epsfxsize=2.0 in\centerline{\epsffile{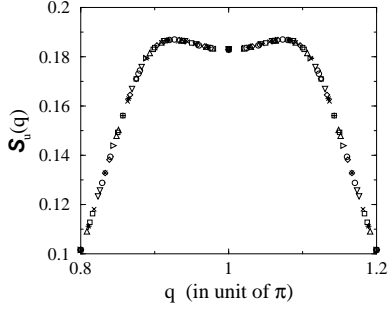}}
\caption[]{ Static structure factor ${\cal S}_u (q)$
for $h_u$=0.5 and $h_s$=0.05. The results for chain
length from 60 to 100 are shown. }
\label{fig5}
\end{figure}

The incommensurate behavior in the correlation function
leads directly to the peak shift from $\pi$ in the static structure factor
${\cal S} (q)$ which can be obtained from the correlation functions.
For the uniform field ($z$) axis and the staggered field ($x$) axis, we can write:
\begin{eqnarray}
{\cal S}_u (q) &=& \frac{1}{L} \sum_l e^{i q l} {\cal C}_u (l),
\label{factoru}
\\
{\cal S}_s (q) &=& \frac{1}{L} \sum_l e^{i q l} {\cal C}_s (l).
\label{factors}
\end{eqnarray}
Since we use the periodic boundary conditions in our calculations,
the  wave vector $q$ is well defined, $q = \frac{2 \pi n}{L}, n=1, \cdots, L$.

In Fig. \ref{fig5}, we present the static structure factor
${\cal S}_u$ at $h_u=0.5$ and $h_s$=0.05 for the even chain 
length from 60 to 100. The results for
different chain length systems fall onto the same curve. 
This success is due to the small correlation length of the system
considered ($\xi \sim 4.05$). The chain lengths used here are
much larger than the correlation length, and the finite size effect is very weak.
Fig. \ref{fig5} shows a two peak structure symmetric about 
$\pi$ which is obtained by Eq. (\ref{factoru}).  
For this case, we have obtained the critical wave vector shift 
$\delta q = |q - \pi| \sim$  0.224 $\sim$ 0.07 $\pi$.
The accuracy of $\delta q$ obtained from the peak deviation of ${\cal S}_u (q)$ 
is good; it is mainly limited by the finite chain lengths used in the
calculations and the error is estimated to be less than 1 \%.

\begin{figure}[ht]
\epsfxsize=3.0 in\centerline{\epsffile{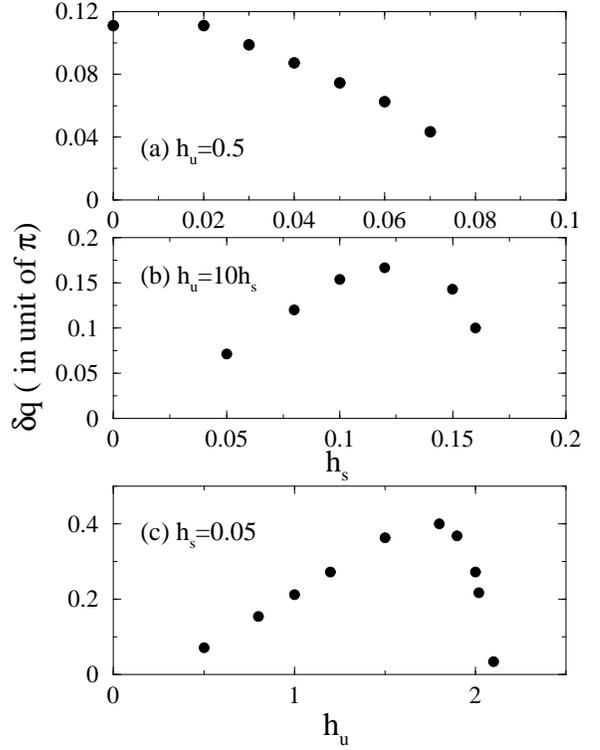}}
\caption[]{ The field dependence of the critical wave vector shift
$\delta q$ for (a) $h_u$=0.5; (b) $h_u = 10 h_s$; and (c) $h_s=0.05$. }
\label{fig6}
\end{figure}

For all  parameters studied in these calculations, 
using the peak position of the static structure factor ${\cal S}_u(q)$ 
to determine the existence of the incommensurate state, 
we found the critical wave vector shift in three
sets of parameters, $h_u=0.5$, $h_u=10 h_s$, and $h_s=0.05$.
The critical wave vector shift $\delta q$ versus   field in these cases
is shown in Fig. \ref{fig6}.
Since the largest total chain length in our calculations is N=100,
we cannot detect the incommensurate state if 
$\delta q < \frac{2 \pi}{100} = 0.02 \pi$.
For the $h_u=0.5$ case, the critical wave vector shift becomes smaller
when the staggered field $h_s$ increases from zero.
It is expected to go to zero when the staggered field is large enough.
In other words, the existence of the staggered field will frustrate the 
incommensurate state.  For the other two cases $h_u = 10 h_s$ and $h_s=0.05$,
the critical wave vector shifts away from $\pi$ with increasing 
uniform field but eventually returns to $\pi$ around $h_u^c$.

For comparison, we examine our calculated results with  parameters corresponding
to the reported experimental case on Cu benzoate. \cite{dender97}
The highest magnetic field reached in the experiment is about
7T which corresponds to $h_u \approx$ 0.52 . 
The energy gap observed at this field
is $\Delta \sim$ 0.4 meV, considering the coupling constant for
the material $J$=1.57 meV, $\Delta /J \sim$ 0.255. From our calculated 
results for $h_u$ = 0.5, we estimate that
the induced staggered field $h_s \sim$ 0.04.  The critical wave vector shift $\delta q$
for $h_u$=0.5, and $h_s$=0.04 is $\delta q \sim$ 0.274 $\sim$ 0.09 $\pi$.
The critical wave vector shift for $h_u$=0.52, $h_s$=0.04 will be slightly
larger than 0.09 $\pi$.  In the experiment, the largest wave shift for H=7T is  
about 0.12 $\pi$.  This comparison shows that the calculated results are
in good agreement with the experiment.

\section{summary}
\label{sec3}

We have carried out systematic calculations using the density matrix renormalization
group method to study the behavior of the energy gap,  magnetization, and 
spin correlation  functions  of   spin-${1 \over 2}$ antiferromagnetic Heisenberg
chain in the presence of a uniform   and a perpendicular
staggered magnetic field. An extensive examination of the parameter space
has revealed many interesting features beyond those reported in previous
studies.  In particular, results at very high magnetic field show
quantitatively  and even qualitatively  different  behaviors in the
energy gap and magnetization,  from those found for
lower fields.  For the $h_u=10 h_s$ case, which is close to the real
parameters in the experiments, we can see from Fig. \ref{fig1} and
Fig. \ref{fig2} that the competition of the staggered and uniform field
is visible only when $h_s > 0.15$, that is $h_u > 1.5$. In Cu benzoate,
$h_u = 1.5$ corresponds to a applied field $H \sim $ 21 Tesla.
Further experimental investigation at  magnetic fields higher than
21 Tesla is needed to test these predictions.

The field-induced energy gap is dominated by the staggered field when the uniform field 
is smaller than the standard spin-${1 \over 2}$ chain saturation field $h_u^c$=2.0.
When the uniform field is comparable or larger than $h_u^c$, the effect of the uniform 
field becomes important and must be taken into account. The uniform field introduces
frustration effects and creates a local minimum in the energy gap near $h_u^c$ in 
several cases. 

The magnetization results clearly reveal the competition of the
uniform  and  staggered fields. 
However,  for some uniform (staggered) field, the existence
of a small staggered (uniform) field enhances
the uniform (staggered) magnetization instead of suppressing it.
The competition of the two kinds of field can also
be seen from the incommensurate behavior with the 
staggered field suppressing the incommensurate state
and moving the critical wave vector closer to the zero-field value $\pi$.

The specific heat experiment
on Cu benzoate\cite{dender97}  shows that the field-induced gap scales
with approximately 2/3 power of the applied magnetic field. 
The analytic results\cite{oshikawa97,affleck99} yield
the same power law dependence for Hamiltonian (\ref{ham}).
In our numerical results, for $h_u =0$ case, the scaling power
$b_L$ =0.678 for the longitudinal gap and $b_T$=0.63 for the transverse gap.
They are in good agreement with the experiment and the analytic value.
The existence of the non-zero uniform field 
modifies the power law relation. When the uniform field is small,
the modification is almost negligible. Further high field experiments 
are needed to examine the predicted effect of the applied uniform field on the
scaling behavior of the induced gap.

\acknowledgments

J. Lou would like to thank Prof. M. Oshikawa, Prof. T. K. Ng, 
Dr. Xiaoqun Wang, and Dr. Tao Xiang for useful discussion. 
This work was supported  by the Department of Energy at the
University of Nevada, Las Vegas and by the Chinese Natural Science Foundation.


\begin{references}

\bibitem{haldane83} F. D. M. Haldane, Phys. Rev. Lett. {\bf 50}, 1153 (1983);
                    F. D. M. Haldane, Phys. Lett. {\bf 93A}, 464 (1983).

\bibitem{affleck89} For a review see Ian Affleck, in {\it Fields, Strings and
                    Critical Phenomena}, edited by E. Br\'ezin and 
                    J. Zinn-Justin (North-Holland, Amsterdam, 1989), p. 511.

\bibitem{dender97}  D. C. Dender, P. R. Hammar, D. H. Reich, C. Broholm,
                    and G. Appli, Phys. Rev. Lett. {\bf 79}, 1750 (1997).

\bibitem{date70}  M. Date, H. Yamazaki, M. Motokawa, and S. Tazawa,
                  Suppl. Prog. Theor. Phys. {\bf 46}, 194 (1970).

\bibitem{dender96}  D. C. Dender, D. Davidovi\'c, D. H. Reich, and C. Broholm,
                    Phys. Rev. B {\bf 53}, 2583 (1996).


\bibitem{muller81} G. M\"{u}ller, H. Thomas, H. Beck, and J.C. Bonner,
        Phys. Rev. B {\bf 24}, 1429 (1981); see also R. Chitra and T. Giamarchi,
        Phys. Rev. B {\bf 55}, 5816 (1997).


\bibitem{oshikawa97} M. Oshikawa and I. Affleck, 
                     Phys. Rev. Lett. {\bf 79}, 2883 (1997).

\bibitem{dm1}  I. Dzyaloshinskii, J. Phys. Chem. Solids {\bf 4}, 241 (1958).

\bibitem{dm2} T. Moriya, Phys. Rev. {\bf 120}, 91 (1960).

\bibitem{affleck99}  I. Affleck and M. Oshikawa, 
                     Phys. Rev. {\bf 60}, 1038 (1999) and refs therein; 
	        			         (Erratum) {\bf 62}, 9200 (2000).

\bibitem{essler98}  F. H. L. Essler and A. M. Tsvelik, 
                    Phys. Rev.  B {\bf 57}, 10592 (1998).


\bibitem{essler99}  F. H. L. Essler, 
                    Phys. Rev. B {\bf 59}, 14376 (1999).

\bibitem{oshikawa99} M. Oshikawa and I. Affleck, Phys. Rev. Lett.
                      {\bf 82}, 5136 (1999).

\bibitem{fled98} A. Fledderjohann, M. Karbach, K.-H. M\"{u}tter,
Eur. Phys. J. B {\bf 5}, 487 (1998); {\it ibid} {\bf 7}, 225 (1999).


\bibitem{dmrg}  S. R. White, Phys. Rev. Lett. {\bf 68}, 3487 (1992);
                Phys. Rev. B {\bf 48}, 10345 (1993).

\bibitem{conf} For a review, see {\it Density-Matrix Renormalization},
               edited by I. Paschel, X. Wang, M. Kaulke, and K. Hallberg,
               Lecture Notes in Physics (Springer, New York, 1999).

\bibitem{spinone} S. R. White and D. A. Huse,  
                  Phys. Rev. B {\bf 48}, 3844 (1993).

\bibitem{spintwo} Shaojin Qin, Xiaoqun Wang, and Lu Yu,
                  Phys. Rev. B {\bf 56}, R14251 (1997).
                  Xiaoqun Wang, Shaojin Qin, and Lu Yu,
                  Phys. Rev. B {\bf 60}, 14529 (1999).

\bibitem{lou99}  Jizhong Lou, Xi Dai, Shaojin Qin, Zhaobin Su, and Lu Yu,
                 Phys. Rev. B {\bf 60}, 52 (1999).

\bibitem{white96} S. R. White and I. Affleck, 
                 Phys. Rev. B {\bf 54}, 9862 (1996).

\end{references}
\end{document}